# Influences of Exciton Diffusion and Exciton-Exciton Annihilation on Photon Emission Statistics of Carbon Nanotubes

Xuedan Ma[1†], Oleskiy Roslyak[1,5], Juan G. Duque[2], Xiaoying Pang[3], Stephen K. Doorn[1], Andrei Piryatinski[4*], David H. Dunlap[6*], and Han Htoon[1*]

[1]Center for Integrated Nanotechnologies, Materials Physics and Applications Division, [2]Physical Chemistry and Applied Spectroscopy, Chemistry Division, [3]High Power Electrodynamics, Accelerator Operations and Technology Division, [4]Theoretical Division, Los Alamos National Laboratory, Los Alamos, New Mexico, 87545, USA. [5]Department of Physics and Engineering Physics, Fordham University, Bronx, New York, 10458, USA. [6]Department of Physics and Astronomy, University of New Mexico, Albuquerque, New Mexico, 87131, USA.

## Abstract

Pump-dependent photoluminescence imaging and 2nd order photon correlation studies have been performed on individual single-walled carbon nanotubes (SWCNTs) at room temperature that enable the extraction of both the exciton diffusion constant and the Auger recombination coefficient. A linear correlation between these is attributed to the effect of environmental disorder in setting the exciton mean free-path and capture-limited Auger recombination at this lengthscale. A suppression of photon antibunching is attributed to creation of multiple spatially non-overlapping excitons in SWCNTs whose diffusion length is shorter than the laser spot size. We conclude that complete antibunching at room temperature requires an enhancement of the exciton-exciton annihilation rate that may become realizable in SWCNTs allowing for strong exciton localization.

Nonclassical photon emission statistics (photon-antibunching) of single quantum emitters such as single atoms [1], molecules [2], quantum dots [3], and nitrogen-vacancy centers in diamond [4] have been recently investigated intensively for development of single photon sources needed for realization of quantum communication . These studies have been extended to single-walled carbon nanotubes (SWCNTs) [5-7], which are near-perfect one-dimensional (1d) semiconductors allowing for fast exciton migration [8-10]. Provided the lowest energy band is populated by more than one exciton, one cannot expect strong photon antibunching from such 1d systems. Studies have revealed that antibunching is possible at cryogenic temperatures due to strong localization of excitons and their subsequent exciton-exciton annihilation (EEA) [5-7], yet so far no systematic studies have been conducted at room temperature on photon statistics of delocalized, 1d excitons. While diffusion, on one hand, facilitates antibunching by allowing for EEA of spatially separated excitons, it could, on the other hand, lead to spreading of the exciton population leading to independent emission of multiple photons. In this Letter, we systematically examine the interplay of these two processes on SWCNT photon emission statistics. This could shed a new light on the feasibility of SWCNT based room-temperature single photon sources.

Exciton diffusion and EEA have been the focii of many previous studies because they play critical roles in defining photoluminescence (PL) characteristics of SWCNTs. Evidence of EEA has been observed in the form of pump-dependent variations in PL decay dynamics [11-13] as well as saturation behaviors of PL [14,15] and transient absorption signals [20]. Many attempts have also been made to determine the exciton diffusion length, $L_D$, in SWCNTs [8,16-19], revealing values ranging from nanometers [9] to hundreds of nanometers [10,16,19]. The large discrepancies in these measurements have been attributed to effects of pump power [10], differences in fabrication methods (CoMoCAT or HiPco) [8,10], and the environment (surfactant- or DNA-wrapped or air-suspended) [8,10] of the SWCNTs. Among these factors, the pump-dependence of $L_D$ arises from an increase of the EEA rate with the square of the exciton population, a consequence of bi-molecular processes [20]. Yet, in spite of the well developed Smolochowski-Noyes theory for reaction diffusion processes [21], the EEA term has not been fully incorporated into the 1d diffusion equation in analysis of the pump dependence of the PL intensity profiles [10]. Our study below reveals that EEA processes in SWCNTs have multiple lengthscale contributions associated with both $L_D$ and the exciton mean free-path, $\xi$.

To address these open issues, we performed pump power-dependent PL imaging and 2$^{nd}$ order photon correlation (g$^{(2)}$) studies on individual (6, 5) SWCNTs. We spread deoxycholate (DOC) wrapped, chirality sorted [22,23] HiPco [24] SWCNTs on glass substrates with density less than one nanotube per 100 µm$^2$ area. A standard micro-PL system (see Supplementary section S1 for details) was used to perform single SWCNT PL imaging and g$^{(2)}$ experiments. Predominantly (6, 5) SWCNTs with lengths distributed in 1.2 to 2.6 µm range were excited at their $E_{22}$ resonance of 570 nm with femtosecond laser pulses. We first imaged each SWCNT by illuminating them entirely using a wide-field mode and subsequently determined their lengths from the FWHM of their axial PL intensity profiles (Fig. 1a and 1g). Then we confocally excited the SWCNTs at their centers with a laser spot having a diameter $\sigma_0 = 455$ nm. Resulting pump-dependent PL images of two SWCNTs with the lengths of 1.2 and 2.6 µm (Figs. 1b-1e and 1h-1k) show PL profiles broadening along their lengths, indicating spreading of the exciton population beyond the extent of the laser spot. For the 1.2 µm SWCNTs (Fig. 1b-1e), the FWHM of the axial PL intensity profile increases with the rise of pump fluence and saturates at very high pump fluences (Fig. 1f). In contrast, in the case of the longer 2.6 µm SWCNTs (Fig. 1h-1k), the profile shrinks with the increase of the pump fluence (Fig. 1l). Data validating our analysis on PL profiles and more examples of pump dependent PL intensity profiles supporting these trends are given in S2-S4 and S5, respectively.

This interesting pump dependence cannot be explained in terms of a linear 1d diffusion equation and requires an explicit nonlinear term describing the EEA processes [10]. Here, we treat EEA as an exciton coalescence reaction (i.e., $A+A \to A$) [25]. For the problem of interest this process is governed by the following equation for the exciton density, $n(x,t)$,

$$\frac{\partial n(x,t)}{\partial t} = -k_{1X} n(x,t) + D \frac{\partial^2 n(x,t)}{\partial x^2} - C_{EEA}(t) n^2(x,t) . \tag{1}$$

Here, $k_{1X}$ is the exciton decay rate determining the PL lifetime in the absence of EEA, $D = L_D^2 k_{1X}$ is the exciton diffusion constant, and

$$C_{EEA}(t) = C_A \left( 1 - \frac{1 - \sqrt{2\varepsilon}\, \text{erf}\left[\sqrt{2\varepsilon \alpha t}\right] - e^{\alpha t(1-2\varepsilon)} \text{erfc}\left[\sqrt{\alpha t}\right]}{1 - 2\varepsilon} \right) \tag{2}$$

is a time-dependent EEA coefficient (dimensionality $nm/ps$) where the prefactor $C_A$ is the Auger recombination coefficient corresponding to the microscopic electron-hole recombination processes at the lengthscale of the exciton mean free-path, $\xi$. In the Smoluchowski-Noyes theory of diffusion-limited reactions leading to Eq. (1), the diffusion of two particles in their center of mass frame is equivalent to the diffusion of a single particle in the presence of a reaction (i.e., recombination) center with twice the diffusion constant. The first term in the parentheses of Eq. (2) (i.e., unity) arises specifically in 1d since a particle in the neighborhood of a reaction center has a high probability of immediately finding it. The second term describes the formation of a depletion region in the vicinity of a reaction center at a rate $\alpha = C_A^2 / 4D$, and ($\text{erfc}[x]$) $\text{erf}[x]$ denotes the (complementary) error function. In S6, we demonstrate that Eq. (2) is a generalization of the constant Auger rate model previously used to interpret EEA signatures observed in ultrafast decay of PL PL [12,26] and transient absorption [12,27] signals, and PL saturation behavior [14,15]. The model interpolates between a constant $C_{EEA} = C_A$ valid at short times, to the time-dependent $C_{EEA}(t) \sim t^{-1/2}$ at long-times used in Refs. [9,28] to interpret the EEA dynamics as 1d diffusion-limited recombination. In addition, this accounts for the limitations on the EEA imposed by the exciton decay process via the dimensionless ratio $\varepsilon = 2k_{1X}/\alpha$ comparing the rate of exciton decay to the rate of depletion region formation.

The goal of the PL data analysis is to attain a single set of diffusion lengths, $L_D = \sqrt{D/k_{1X}}$ ($k_{1X} = 1/55$ ps$^{-1}$ for DOC-wrapped SWCNTs [16]) and Auger coefficients, $C_A$, fitting *all* the intensity profiles *for each tube* measured at *different* excitation powers. For this purpose a fitting procedure described in S7 and based on Eqs. (1) and (2) is used. In contrast to prior studies [13, 21, 24] where single intensity profiles were fit to extract $L_D$, we have simultaneously fit 4 to 9 intensity profiles composed of 80 to 180 data points [25] and extracted a *single* set of $L_D$ and $C_A$ for each SWCNT. Fig. 1f and 1l display examples of the best-fit curves along with the experimental data points providing $L_D = 289$ nm, $C_A = 382$ nm$^2$/ps and $L_D = 718$ nm, $C_A = 447$ nm$^2$/ps, respectively. Notice that in Figs. 1f the inset represents a general trend for the widths, $\sigma$, of *all* measured PL profiles to *broaden* with the increasing pump power provided SWCTNs have $L_D < \sigma_0$. In contrast, inset in Fig. 1l shows that SWCNTs having $L_D > \sigma_0$ demonstrate *narrowing* of the PL profile with increasing pump power. Numerical simulations discussed in S8 and based on Eqs. (1) and (2) clearly reproduce these trends.

To rationalize the trends, we estimate average PL profile widths as $\sigma^2 = \sigma_0^2 + \Delta\sigma^2$, where $\Delta\sigma^2 = 2D\tau_{EEA}$ describes the contribution of exciton diffusion to the profile spread on the EEA timescale $\tau_{EEA} \simeq \dfrac{D\sigma_0}{(N_0 - 1)C_A^2}$. Here, $N_0$ is the number of excitons prepared by the pulse that can be evaluated from absorption cross section of SWCNT ($\bar{\sigma} = 10^{-12}$ cm$^2$/μm) as shown in S7. [29,30] This timescale is estimated by solving Eq. (1) describing EEA at short times by setting $k_{1x} = 0$ as described in S6. Thus, $\Delta\sigma^2 \sim 1/N_0$ indicating that the PL profiles should narrow if the number of excitons (pump power) increases. Since the natural upper limit on $\Delta\sigma$ is $L_D$, one can expect this trend to be observed for the initial profile widths $\sigma_0 < L_D$. This is exactly the case in Fig. 1l. In the opposite case ($\sigma_0 > L_D$) the diffusive contributions to the PL profiles become negligible. The EEA processes occur predominately in the center of the Gaussian pulse where the exciton concentration is the highest. This reduces the amount of PL intensity relative to the radiation of the Gaussian tails, effectively leading to a spatial broadening of the integrated radiation profile, (Fig. 1f inset). In the same wide-profile limit, it should also be noted that two excitons initially separated by a distance $\gtrsim L_D$ never interact over the course of the experiment and effectively belong to different segments of a SWCNT and can be considered to be independent emitters. An exciton density prepared within the laser spot of width $\sigma_0 \gg L_D$ can be split into $m \approx \sigma_0 / L_D$ independent emitters.

Fig. 2 presents values of $C_A$ vs $D = L_D^2 k_{1X}$ obtained from a global fit of the PL intensity profiles of 26 SWCNTs, yielding diffusion constants in the range of $D \sim 6 \times 10^2 - 4.4 \times 10^3$ nm$^2$/ps ($L_D \sim 182 - 492$ nm). This is in good agreement with the results of recent measurements [8,16]. At the same time, our analysis yields Auger coefficients in the range $C_A \sim 200 - 750$ nm/ps. Although $C_A$ and $D$ vary for each SWCNT, we find a strong *linear* correlation between them.

To rationalize the correlation between $C_A$ and $D$, we recall that the exciton energy landscape in SWCNTs is perturbed by environmentally induced disorder[16]. This limits exciton coherent motion characterized by mean velocity, $v$, to a mean free-path, $\xi$, giving rise to a diffusion constant $D = v\xi$. By taking into account that $v = \sqrt{k_B T / m^*}$, where $k_B T$ is thermal energy and the ratio of exciton effective mass to electron mass in vacuum is $m^*/m_0 \sim 0.05$, we use experimental values in Fig. 2, to estimate the variation range of mean free-path to be $\xi \sim 4-15$ nm. We further assume that two-excitons form *a bound state* with a delocalization length *less* than $\xi$ before experiencing Auger recombination. As demonstrated in S9, in contrast to direct Auger recombination [31], such a *capture-limited* Auger process can be described by the Auger constant being proportional to $\xi$. Specifically, $C_A = \xi / 2\tau_{2X,A}$, where, $\tau_{2X,A}$ is native Auger recombination time of the bound two-exciton complex. Therefore, observed variation in the values of $D$ and $C_A$ reflects an environment-induced variation of $\xi$ in the ensemble of SWCNTs. By taking the ratio of $D$ to $C_A$ we find the experimentally observed linear correlation whose slope value gives $\tau_{2X,A} = 10$ fs (see S9). Furthermore, capture-limited Auger recombination requires formation of a bound two-exciton state on the timescale less than $\tau_{2X,A}$. The analysis in S9 suggests that such an ultrafast exciton capture might be facilitated by Coulomb mediated long-range exciton-exciton interactions.

Crossover between single and multiple independent quantum emitter regimes in SWCNTs has clear signatures in photon emission statistics. The quantity of interest is the degree of photon-antibunching $R_0$ at the low pump fluence limit available from the measurements of 2nd order photon correlation function, $g^{(2)}$. For *a single* quantum emitter in the small spot size limit ($\sigma_0 < L_D$), $R_0 = Q_{2X}/Q_{1X}$ where $Q_{2X}$, and $Q_{1X}$ are two-exciton-to-exciton and exciton quantum yields, respectively. [32,33] In the limit of large spot size ($\sigma_0 > L_D$), the *minimum* attainable degree of photon-antibunching is $R_0 = (m-1)/m$. Here, $m = \sigma_0/(cL_D)$ is the number of independent emitters each of size $cL_D$ where a scaling parameter $c \sim 1-2$ is introduced to facilitate fitting of the experimental data. We use a generalized expression for $R_0$ accounting for $m$ (including $m=1$) independent emitters and two exciton emission/recombination processes given by [34]

$$R_0 = \frac{p_2 Q_{2X}}{m Q_{1X}} + \frac{m-1}{m}. \qquad (3)$$

Here, $p_2 = 2P_2^{\langle N \rangle}/\left(P_1^{\langle N \rangle}\right)^2$ is the ratio of Poisson probabilities to produce two and one excitons, respectively, and $\langle N \rangle$ is the average number of absorbed photons.[35]

To examine the significance of the two terms in Eq. (3), we performed pump-dependent $g^{(2)}$ measurements on 15 SWCNTs and analyzed $R_o$ in terms of the ratio $\sigma_0/L_D \sim m$ for fixed $\sigma_0 = 455$ nm, using values of $L_D$ obtained from the analysis of PL intensity profiles. Figs. 3a-3c display representative pump-dependent $g^{(2)}$ traces of 3 SWCNTs with the shortest, medium, and the longest values of $L_D$ having $\sigma_0/L_D = 2.5, 1.4$, and 0.8. In Fig. 3d, $R_0$ is plotted as a function of pump fluence expressed in terms of the average number of absorbed photons per pulse, $\langle N \rangle$ (see S7). The data clearly show that the $R_0$ values gradually increase with $\langle N \rangle$ for all SWCNTs. This trend is more pronounced for the medium-$L_D$ SWCNTs (Fig. 3d) showing the smallest the smallest value of $R_0 = 0.4$ (Fig. 3b, bottom trace) at the lowest pump fluence. However, for the longest and shortest diffusion length SWCNTs (bottom traces in Fig. 3a and 3c), we observe $R_0 > 0.6$.

The $R_0$ values vs. $\sigma_0 / L_D$ measured from all 15 SWCNTs are plotted in Fig. 4. This plot shows that while minimum values of $0.4 \leq R_0 \leq 0.5$ are observed in SWCNTs with $L_D / \sigma_0 = 1.0 - 1.7$, $R_0$ increases for both smaller and larger values of $\sigma_0 / L_D$. To interpret the observed trends, we have evaluated $R_0$ using Eq. (3) along with an independent quantum emitter model (see S10) parameterized by the values of $L_D$ and $C_A$ attained from the analysis of the PL intensity profiles. Calculated total $R_0$ and contributions due to each term in Eq. (3) are shown in Fig. 4. A best fit is obtained by setting the size of each independent emitter to $cL_D = 1.6 L_D$ and the prefactor $p_2 = 1.5$ corresponding to $\langle N \rangle = 0.4$ (consistent with Fig. 3c). For $\sigma_0 / L_D \gtrsim 1.6$ ($m \gtrsim 1$), the calculation shows excellent agreement with experiment and is dominated by the multiple emitter term $(m-1)/m$. As the diffusion length approaches the laser spot size ($\sigma_0 / L_D \sim 1.2 - 1.6$), the contribution of $(m-1)/m$ decreases toward zero approaching the single quantum emitter regime ($m \sim 1$) in which the ratio $Q_{2X} / Q_{1X}$ fully determines $R_0$. Constant behavior of $R_0$ at this length scale is due to the weak dependence of the EEA rate on $L_D$ illustrated in S10.[36] For $\sigma_0 / L_D < 1$, the calculation of $R_0$ does not reproduce the experimental trend, revealing the limitation of the coarse-grained estimate of $Q_{2X}$ (see S10).

Observed limitations for $R \gtrsim 0.4$ seem to contradict experimental reports of strong antibunching found in SWCNT low temperature PL emission [5-7]. However, at cryogenic temperatures, potential fluctuations along the length of the SWCNT become deep enough to trap the excitons. These traps become recombination centers for all the excitons created within $L_D$ and yield highly localized PL emission regardless of $L_D$. Since the exciton trapping strongly enhances the efficiency of the EEA process, PL emission of a single trap site exhibits near complete antibunching. This type of localization-induced antibunching cannot be observed for band-edge excitons at room temperature due to large thermal fluctuations. However, it may become possible in oxygen doped SWCNTs that emit from a deep trap level located 160 meV below the band edge exciton [37].

In conclusion, the pump-dependent PL intensity profiles measured from individual SWCNTs have been examined. An analysis based on the Smolochowski-Noyes reaction-diffusion equation with a time-dependent EEA coefficient reveals a linear correlation between $C_A$ and $D$ that is attributed to the effect of environmental disorder in setting the exciton mean free-path whose values are estimated to vary in the range of $\xi \sim 4-15$ nm. Linear dependence of the Auger coefficient on mean free-path, $C_A \sim \xi$, further suggests ultrafast capture-limited Auger recombination of exciton pairs. Both pump-dependence of PL profile widths and 2$^{nd}$ order photon correlation measurements confirm creation of multiple spatially non-overlapping excitons forming independent emitters in SWCNTs whose diffusion length is shorter than the laser spot size, i.e., $\sigma_0 \gtrsim L_D$. The 2$^{nd}$ order photon correlation measurements provide a value of $1.6 L_D$ for the size of an independent quantum emitter, and clearly show a drop in the minimal degree of photon antibunching to $R_0 \sim 0.5$ as the number of independent emitters approaches unity. Further reduction of $R_0$ requires significant decrease in the two-exciton-to-exciton quantum yield, $Q_{2X}$. This can be achieved at room temperature via an enhancement of the EEA rate and may become realizable in SWCNTs with strong exciton localization. These findings have pertinence for the development of SWCNT-based room-temperature single photon sources, and the analysis may be applied to other 1d nanomaterials such as conjugated polymers and semiconductor nanowires.

This work was conducted, in part, at the Center for Integrated Nanotechnologies (CINT), a U.S. Department of Energy, Office of Basic Energy Sciences (OBES) user facility and supported in part by Los Alamos National Laboratory Directed Research and Development Funds.
This work was conducted, in part, at the Center for Integrated Nanotechnologies (CINT), a U.S. Department of Energy, Office of Basic Energy Sciences (OBES) user facility and supported in part by Los Alamos National Laboratory Directed Research and Development Funds.

*Corresponding authors:
AP: apiryat@lanl.gov
DHD: dunalp@unm.edu
HH: hhtoon@lanl.gov

†Present address: Center for Integrated Nanotechnologies, Sandia National Laboratories, Albuquerque, New Mexico 87185, USA

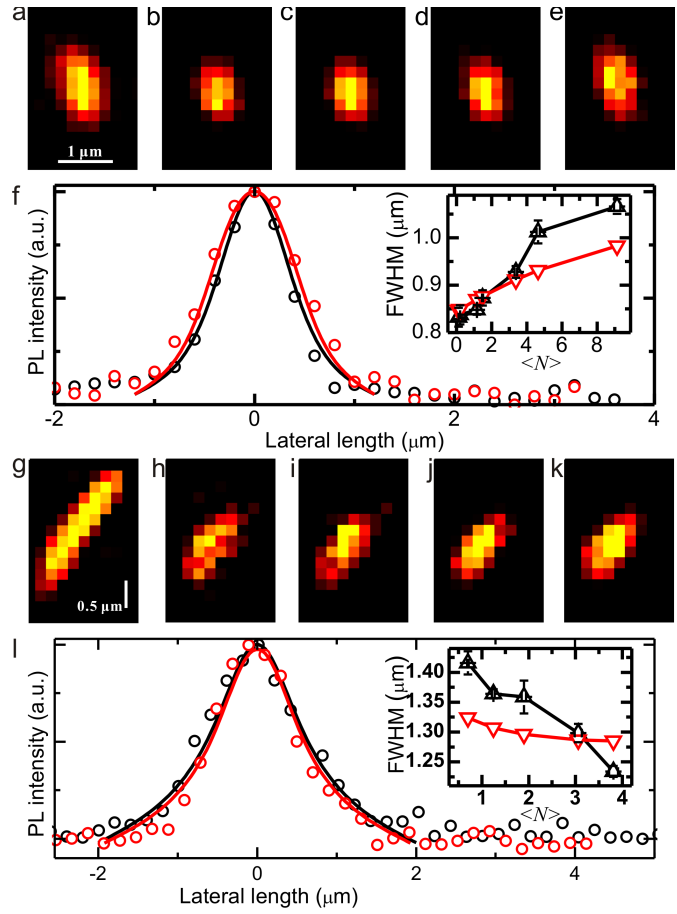

**FIG. 1: a, g**, Wide-field PL images of two SWCNTs with lengths of ~ 1.2 µm (a) and 2.6 µm (g), respectively. **b-e & h-k**, PL images of the same SWCNTs confocally excited at the center of the tubes. The excitation intensity was increased from 6.6 W/cm$^2$ (b) to 535.0 W/cm$^2$ (e), and 41.2 W/cm$^2$ (h) to 223.1 W/cm$^2$ (k), respectively. **f, l,** Axial intensity profiles of SWCNT PL images (black circles: b & h; red circles: e & k) together with the corresponding fitting curves (solid lines). **Insets,** FWHM of measured intensity profiles (black triangles), uncertainty (error-bars) determined from Gaussian fit of the intensity profiles (see S3) and FWHM of calculated (red triangles) PL intensity profiles are plotted as the function of pump fluence given in terms of average number of absorbed photons per laser pulse $\langle N \rangle$.

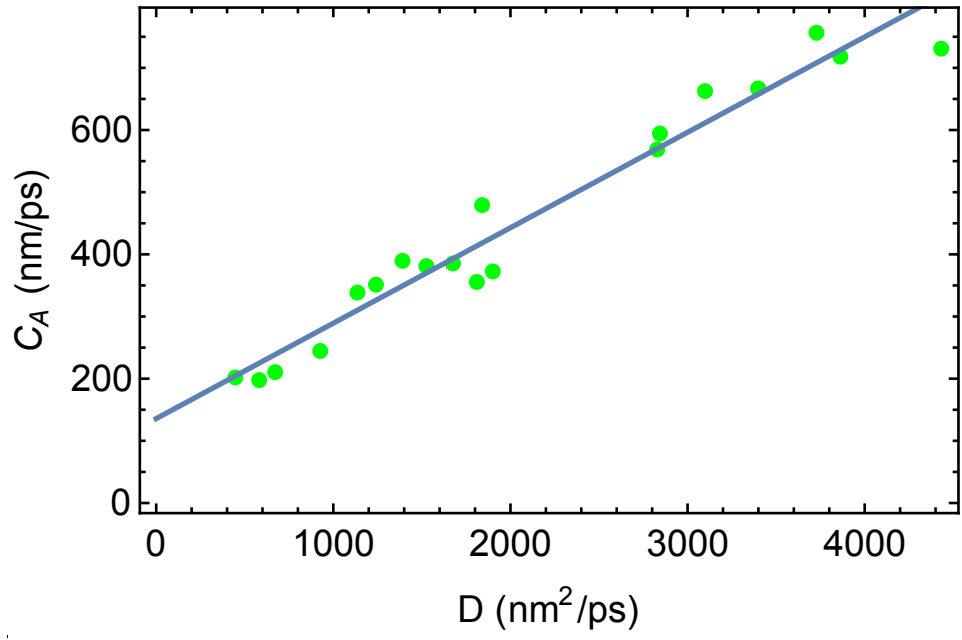

**FIG. 2:** Dots show values of Auger coefficient $C_A$ and diffusion constant, $D$, obtained from global fitting of the intensity profiles of individual SWCNTs using our generalized diffusion model (Eqs. (1) and (2)). Straight line $C_A \sim D/d_0$ with parameter $d_0 = 6.5$ nm is a linear fit of the experimental points.

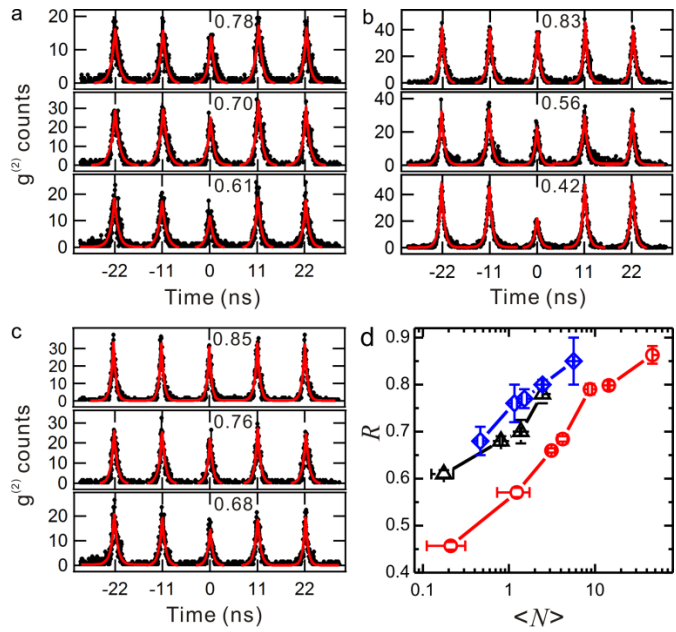

**FIG. 3: a-c**, Representative $g^{(2)}(\tau)$ of three individual SWCNTs with diffusion lengths of 188 nm (a), 336 nm (b), and 525 nm (c) measured at increasing pump fluences (from bottom to top, black dots). Each peak is well fitted by a two-sided exponential function (red curves). The degree of photon-antibunching, $R$, is indicated for each curve. **d**, $R$ of the three individual SWCNTs in part a-c plotted as a function of $<N>$ (black: a; red: b; blue: c).

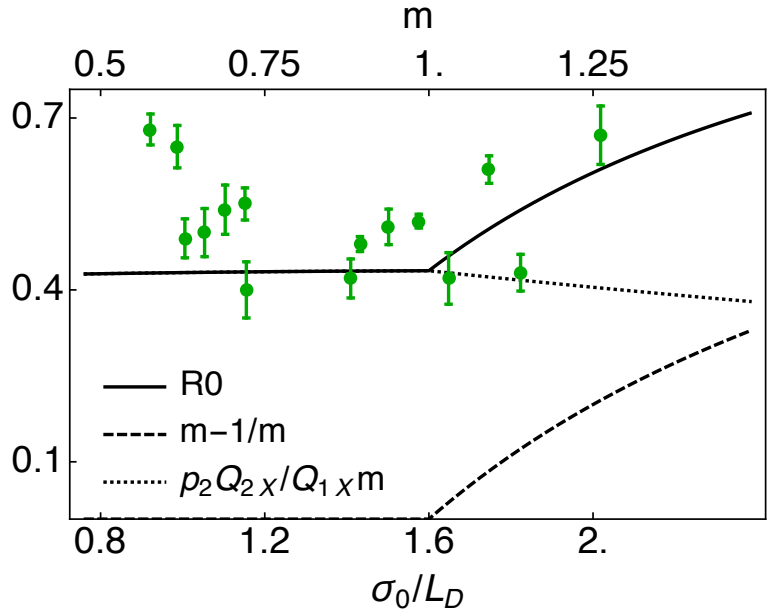

**FIG. 4:** Measured (green circles) and calculated (solid line) $R_0$ values plotted along with calculated $p_2Q_{2X}/(mQ_{1X})$ (dotted line) and $(m-1)/m$ (dashed line) factors as a function of the ratio $\sigma_0/L_D$ and number of independent emitters $m = \sigma_0/1.6L_D$.

# Supplementary Information:

# Influences of Exciton Diffusion and Exciton-Exciton Annihilation on Photon Emission Statistics of Carbon Nanotubes

Xuedan Ma[1], Oleksiy Roslyak[1,5], Juan G. Duque[2], Xiaoying Pang[3], Stephen K. Doorn[1], Andrei Piryatinski[4*], David H. Dunlap[6*], and Han Htoon[1*]

[1]Center for Integrated Nanotechnologies, Materials Physics and Applications Division, [2]Physical Chemistry and Applied Spectroscopy, Chemistry Division, [3]High Power Electrodynamics, Accelerator Operations and Technology Division, [4]Theoretical Division, Los Alamos National Laboratory, Los Alamos, New Mexico, 87545, USA. [5]Department of Physics and Engineering Physics, Fordham University, Bronx, New York, 10458, USA. [6]Department of Physics and Astronomy, University of New Mexico, Albuquerque, New Mexico, 87131, USA.

*This document contains the following supplementary information:*

**S1 Experimental methods.**
**S2 Laser spot size at different excitation powers.**
**S3 FWHMs of PL intensity profiles and their uncertainties.**
**S4 PL stability of SWCNTs.**
**S5 Pump dependence of PL intensity profiles.**
**S6 Limiting cases of diffusion equation accounting for EEA dynamics.**
**S7 Analysis of the PL intensity profiles using a generalized 1d diffusion equation.**
**S8 Numerical simulation of PL intensity profile width vs exciton population.**
**S9 Auger recombination model.**
**S10 Independent quantum emitter and coarse-grained EEA models.**

**S1 Experimental methods**

**Sample preparation**. SWCNTs used in this study were synthesized by the HiPco method [24]. Colloidal SWCNTs suspended in 1% aqueous deoxycholate were obtained by shear mixing for 1h. Large bundles and insoluble materials were then removed by centrifugation. The resulting supernatant was chirality-sorted by nonlinear density gradient ultracentrifugation [22,23]. Suspension fractions enriched in (6,5) tubes were collected and stored for PL measurements. We restricted our experiments to a single chirality in order to avoid any chirality dependent effects.

To prepare samples for individual SWCNT PL measurements, a 5 – 10 μL drop of nanotube suspension was spread on a pre-cleaned, plasma-treated glass cover slip. To reduce water evaporation and Brownian motion of the SWCNTs, a second smaller cover slip sealed on three sides with vacuum grease was used to spread and cover the drop. With this procedure, the surfactant layer on the SWCNT surface remained largely intact and PL of SWNTs remained stable.

**Photoluminescence imaging and photon correlation spectroscopy**. PL measurements were performed with a home-built confocal laser microscope. A mode-locked Ti:sapphire laser (Chameleon, Coherent) together with an optical parametric oscillator (APE, Coherent) was used to selectively excite (6,5) SWCNTs at their second allowed optical transition ($E_{22}$) with 571 nm, 130 fs laser pulses at 90 MHz repetition rate. The excitation laser beam was focused by an oil immersion objective (100x, NA=1.30, Olympus). The emitted light from the sample was collected by the same objective, filtered by a 890 nm long-pass filter (Chroma), and directed to a charge-couple device (CCD) camera (Princeton Instrument) with a 1/3 meter spectrograph (Acton 2300i) for imaging and spectroscopy experiments. For each studied tube, pump power dependent PL images were acquired by alternatively increasing and decreasing excitation power to ensure that no hysteresis in the PL intensity was observed (see Supplementary section S4, Fig. S4a). Any tube showing aging during this process was excluded from further measurements. For photon correlation measurements, emission from the SWCNTs was split by a 50/50 beam splitter and sent to two single-photon avalanche diodes (SPAD, PerkinElmer) in the Hanbury-Brown and Twiss arrangement. A hydra-harp time correlated single photon counting module from PicoQuant was used to acquire $g^{(2)}$ functions and PL time traces. At the lowest pump fluence, an integration time of up to 17 hours was used to attain a $g^{(2)}$ trace with reliable signal-to-noise ratio. All the SWCNTs analyzed in Fig. 4 in the main text exhibit very stable PL intensities over the duration of the entire experiment (See Supplementary section S4, Fig. S4b). Any SWCNT that showed rapid PL fluctuation and/or photo-bleaching was excluded from the analysis. For wide-field imaging, an achromatic lens was inserted to focus the incoming collimated laser beam to the back-aperture of the objective. The out coming laser beam had a spot size of ~ 60 μm in diameter.

## S2 Laser spot size at different excitation powers

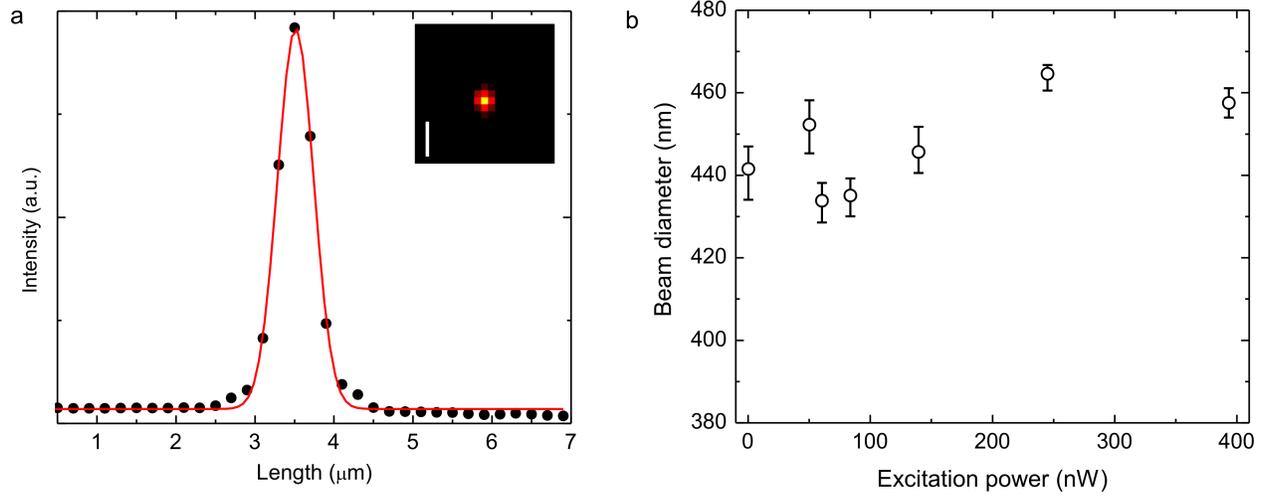

**FIG. S2: a**, A representative intensity profile of a focused laser spot (black dots) together with a Gaussian fitting curve (red curve). The beam diameter $\sigma_0$ of the Gaussian function is ~ 445 nm. Inset: An image of a focused laser spot. The scale bar is 1 μm. **b**, Diameters of laser spot at different excitation powers showing little fluctuation.

## S3 FWHMs of PL intensity profiles and their uncertainties

We fit experimentally measured intensity profiles with Gaussian functions using the least-squares fit algorithm of the Origin graphics program. The fits yield FWHMs of (a) 0.827±0.024 μm for black data points of Fig. 1f; (b) 1.081±0.025 μm for red data points of Fig. 1f; (c) 1.195±0.032 μm for red data points of Fig. 1l and (d) 1.452±0.039 μm for black data points of Fig. 1l, respectively (Fig. S3). These values demonstrate that our data have sensitivity to determine the FWHM with <3% error (i.e. 0.024/0.827). In our pump intensity-dependent measurements, we reject all the low pump power PL intensity profiles that do not allow determination of FWHM with less than 5% uncertainty. For this experiment, we acquired a total of ~290 PL intensity profiles of 43 SWCNTs and 60 intensity profiles were rejected for having insufficient signal to noise ratios.

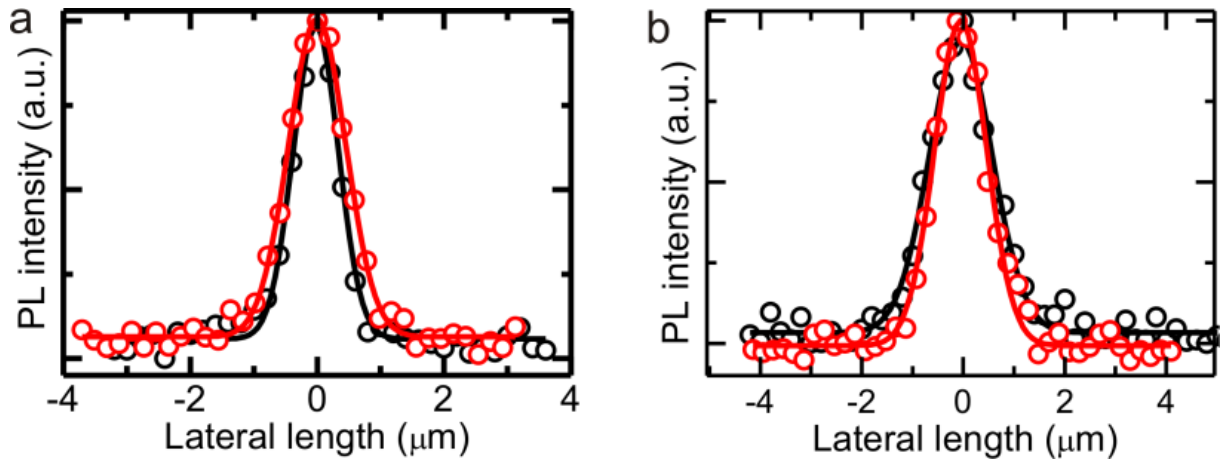

**FIG. S3.** PL intensity profiles shown in Fig. 1f and 1l (main text) fitted by Gaussian functions. The fitting yield FWHM values of 0.827±0.024 μm (black, Fig. 1a), 1.081±0.025 μm (red, Fig. 1a), 1.452±0.039 μm (black, Fig. 1b), and 1.195±0.032 μm (red, Fig. 1b).

**S4 PL stability of SWCNTs**

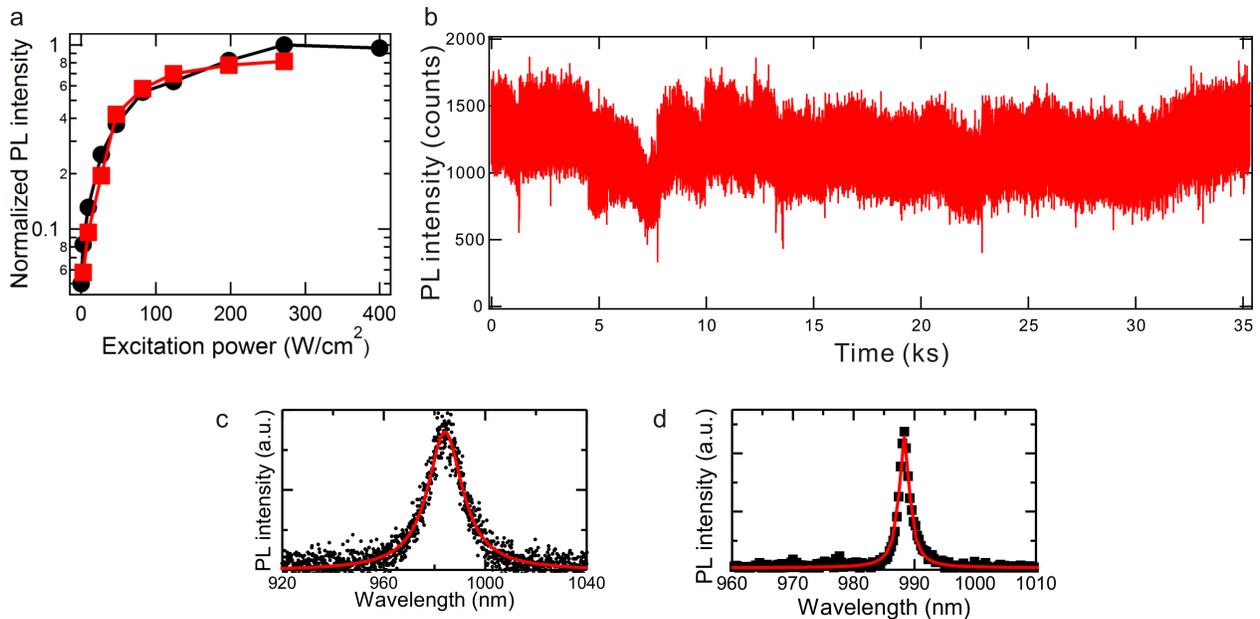

**FIG. S4:** **a**, PL intensity of a SWCNT during successive increasing (black dots) and decreasing (red squares) excitation power cycles plotted on a log-linear scale. The data show no hysteresis associated with photo-aging. **b**, A PL time trace of a SWCNT recorded during ~ 9 hour-long $g^{(2)}$ measurement at ~7W/cm$^2$. The trace shows no significant PL blinking or bleaching. **c**, **d**, Representative PL spectra of SWCNTs measured at room temperature (c) and 5K (d). Red curves are Lorentzian fittings. Narrow, single-peaked spectra indicate individual SWCNTs were studied.

## S5 Pump dependence of PL intensity profiles

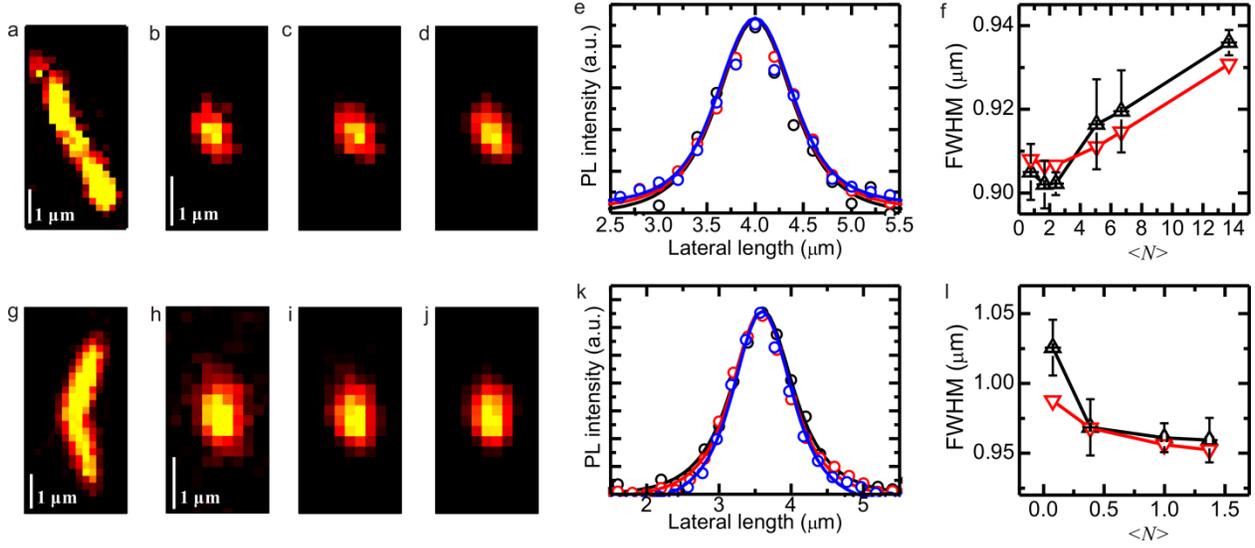

**FIG. S5: a, g**, Wide-field PL images of two SWCNTs with lengths of ~ 2.9 µm (a) and 2.5 µm (g), respectively. **b-d & h-j**, PL images of the same SWCNTs confocally excited at the center of the tubes. The excitation intensity increases from 140 W/cm$^2$ (b) to 802 W/cm$^2$ (d), and 22 W/cm$^2$ (h) to 80 W/cm$^2$ (j), respectively. **e, k**, Axial intensity profiles of confocally excited PL images (black circles: b & h; red circles: c & i; blue circles: d & j) together with the corresponding fitting curves (solid lines). **f, l**, FWHMs of experimentally measured (black triangles) and calculated (red triangles) PL intensity profiles plotted as the function of pump fluence given in terms of averaged exciton occupancy ⟨N⟩.

## S6 Limiting cases of diffusion equation accounting for EEA dynamics

As discussed in the Letter, the timescale for the depletion layer evolution is determined by the rate $\alpha = C_A^2/D$ where $C_A$ and $D$ are the Auger coefficient and diffusion constant, respectively. The dimensionless ratio $\varepsilon = 2k_{1X}/\alpha$ allows us to compare the contribution of the depletion layer evolution on the timescale of time-integrated PL measurements defined by the exciton decay rate $k_{1X}$. Accordingly, we consider the following limiting cases.

**(i) Short time limit.** If an experiment probes short times such as $\alpha t \ll 1$ then the time-dependent EEA rate (C.f. Eq. (2) in the main text) becomes

$$C_{EEA}(t) \approx C_A\left(1 - C_A\sqrt{\frac{t}{\pi D}}\right). \tag{S6.1}$$

Neglecting weak time dependence, we arrive at the following diffusion equation,

$$\frac{\partial n(x,t)}{\partial t} = D\frac{\partial^2 n(x,t)}{\partial x^2} - k_{1X}n(x,t) - C_A n^2(x,t). \tag{S6.2}$$

This constant rate model has been previously used to interpret EEA signatures observed in ultrafast decay of PL [12,26] and transient absorption [12,27] signals, and PL saturation behavior [14,15].

**(ii) Long time limit no PL decay.** In the *absence* of the exciton decay, i.e. $k_{1x} = 0$, the long time $\alpha t \gg 1$ approximation of the main text Eq. (2) results in the EEA rate

$$C_{EEA}(t) \approx 2\sqrt{\frac{D}{\pi t}} \ .\tag{S6.3}$$

Introducing this rate into the diffusion equation, one obtains

$$\frac{\partial n(x,t)}{\partial t} = D\frac{\partial^2 n(x,t)}{\partial x^2} - 2\sqrt{\frac{D}{\pi t}}n^2(x,t) .\tag{S6.4}$$

This expression does not depend on the Auger coefficient and describes a *1d diffusion-limited* EEA process characterized by persistent expansion of the depletion layer and characterized by a long-time $t^{-1/2}$ behavior arising from the restrictions of 1-d topology; only in 3d are there sufficient spatial pathways to replenish the depletion region so that it reaches steady state, leading to the well-known constant Smoluchowski rate. Equations with $C_{EEA}(t) \sim t^{-1/2}$ were used in Refs. [9,28] to interpret the EEA dynamics.

**(iii) Long time limit with PL decay.** If an experiment probes long times, i.e. $\alpha t \gg 1$, at *finite* exciton decay times $\tau_{1X} = 1/k_{1X}$ then the EEA coefficient becomes

$$C_{EEA} \approx C_A \frac{\sqrt{2\varepsilon}}{1+\sqrt{2\varepsilon}} = \frac{2\sqrt{k_{1X}D}}{1+2\sqrt{k_{1X}D}/C_A} .\tag{S6.5}$$

Note that in the time-integrated experiment the long-time behavior is important provided the exciton decay rate is much slower than the depletion layer evolution rate, i.e., $\varepsilon \ll 1$. In this limit the expression above may be approximated as

$$C_{EEA} \approx 2\sqrt{k_{1X}D} \ .\tag{S6.6}$$

Note that substitution of $t = \tau_{1X}/\pi = 1/\pi k_{1X}$ in Eq. (S6.3) exactly recovers Eq. (S6.6), indicating that the exciton decay time determines the limit for the depletion layer growth in the limiting case (ii).

**S7 Analysis of the PL intensity profiles using a generalized 1d diffusion equation**

In order to determine the Auger coefficient, $C_A$, and the diffusion constant, $D$, from the PL spatial profile, we employ the following global fitting algorithm. For a given SWCNT confocally excited at the center, we define a set of measured PL intensity profiles along the tube as $I(x_i, N_0)$, where $x_i$ is a finite set of lateral coordinates and

$$N_0 = \frac{2\bar{\sigma}P_0}{\pi\sigma_0 f\hbar\omega}\tag{S7.1}$$

is the number of excitons prepared by a pump pulse of powers $P_0$. Here $\sigma_0$ is the spatial width of the Gaussian pump, $\hbar\omega$ is the photon energy, and $f$ is the laser pulses repetition rate. The value of absorption cross-section $\bar{\sigma} = 10^{-12}\ cm^2$ has recently been obtained via single nanotube microscopy [29,30]. The background effects are eliminated by a downshift such that the spline-interpolated function, $I(x;N_0)$, peaks at the nanotube center and vanishes at its ends. The measured PL intensity $I(x;N_0)$ is related to the calculated exciton density $n(x,t;N_0,D,C_A)$ as $I(x;N_0) = B \int_0^{T_{max}} n(x,t;N_0,D,C_A)dt$, where the proportionality constant $B$ is defined by the photon collection efficiency. The exciton density $n(x,t;N_0,D,C_A)$ is obtained via numerical solution of Eqs. (1) and (2) in the main text with the boundary conditions $n(x=\pm l/2, t) = 0$, and the initial condition written as a series,

$$n(x,0) = \frac{N_0}{\sqrt{\pi \sigma_0^2/2}} \left\{ e^{-2x^2/\sigma_0^2} + \sum_{n=1}^{M}(-1)^n \left( e^{-2(x-nl)^2/\sigma_0^2} + e^{-2(x+nl)^2/\sigma_0^2} \right) \right\}, \quad (S7.2)$$

This slightly-modified Gaussian initial condition uses the method of images to impose vanishing of n(x,0) at the boundaries since this must be the case in experiment. Here $l$ is the SWCNT length.

To determine the values of D and $C_A$, we have fit the normalized PL,

$$\langle n(x;N_0,D,C_A) \rangle_x = \frac{n(x;N_0,D,C_A)}{\int_{-l/2}^{l/2} n(x;N_0,D,C_A)dx} \approx \frac{I(x;N_0)}{\int_{-l/2}^{l/2} I(x;N_0)dx} = \langle I(x;N_0) \rangle_x, \quad (S7.3)$$

to the numerical solution of Eq. (1). The global fit searches for the minimum of the cumulative error function defined as

$$Er(D,C_A) = \sum_{N_0} \int_{-l/2}^{l/2} dx \left[ \langle n(x;N_0,D,C_A) \rangle_x - \langle I(x,N_0) \rangle \right]^2 \quad (S7.4)$$

The relative error for the results shown in Fig. 2 of the main text is constrained so that $Er < 10^{-4}$.

The set of $(D,C_A)$ values obtained from the ensemble of nanotubes shows a correlation between $C_A$ and D, and can be well fit with the linear function,

$$C_A = C_{A,0} + \frac{D}{d_0}, \quad (S7.5)$$

whose parameters are $C_{A,0} = 136$ nm/ps and $d_0 = 6.5$ nm. Furthermore, Fig S7a, presents extended version of Fig. 2 from the main text superimposed with contour plots of the dimensionless ratio $\varepsilon = 2k_{1X}/\alpha$. According to S6, this ratio allows for comparison of the contribution of the exciton depletion layer evolution on the timescale of time-integrated PL measurements defined by the exciton decay rate $k_{1X}$. The plot shows that the experimental data belong to the region characterized by $\varepsilon \ll 1$. This is a clear indication that the depletion layer evolution has strong contribution to the exciton dynamics on the scale of time-integrated PL measurements. This necessitates the use of the generalized time-dependent Auger coefficient introduced by Eq. (2) as opposed to the often used short time approximation described in S6(i).

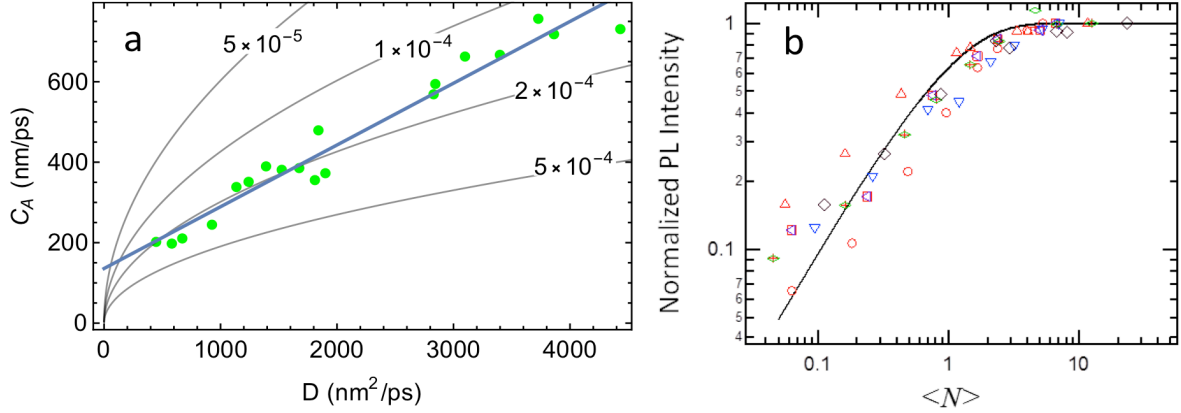

**FIG. S7 a:** Extended version of the main text Fig. 2. Dots show values of Auger coefficient, $C_A$, and diffusion constant, $D$, obtained from global fitting of the intensity profiles of individual SWCNTs using our generalized diffusion equation. The straight line is a data point fit with Eq. (S7.5). Grey lines show contour plot of the dimensionless parameter $\varepsilon = 2k_{1X}/\alpha$ with $\alpha = C_A^2/D$. **b:** Normalized PL intensity saturation curves of 8 individual SWCNTs plotted as the function of $\langle N \rangle$ calculated using $\bar{\sigma} = 10^{-12}$ cm$^2$/μm. Solid line: Theoretical PL saturation curve.

Although almost all the parameters required to estimate $N_0$ (i.e. $P_0, \sigma_0$, and $f$) can be measured accurately, the absorption cross-section of SWCNT must be chosen; we have set $\bar{\sigma} = 10^{-12}$ cm$^2$/μm, a value reported in Ref. [29]. In practice, the value of $\bar{\sigma}$ could vary significantly from one individual SWCNT to another. To check the accuracy of this value, we analyze the PL intensity saturation behavior of individual SWCNTs. In Fig. S7b, we plot the normalized PL intensity saturation curves (normalized to 1 at the saturation PL intensity) of 8 different individual SWCNTs as the function of $\langle N \rangle$ calculated using $\bar{\sigma} = 10^{-12}$ cm$^2$/μm. Clustering of the data points of each of the different individual SWCNTs around the theoretical population saturation of a two-level system (i.e. $(1-e^{-\langle N \rangle})$, solid curve in Fig. S7.2 provides a clear indication that $\bar{\sigma} = 10^{-12}$ cm$^2$/μm is a reasonable estimate. However, the spread of the data points along the $\langle N \rangle$ axis suggests that the actual $\bar{\sigma}$ values of some individual SWCNTs could differ from this average value.

### S8 Numerical simulation of PL intensity profile width vs exciton population

To understand how the FWHM of PL intensity profiles evolve with pump powers, we solve Eqs. (1) and (2) from the main text numerically, taking into account the correlation between Auger coefficient and diffusion constant (diffusion length) given by Eq. (S7.5). The exciton density, $n(x,t)$, is evaluated for a 4 μm long SWCNT with diffusion lengths $L_D$ = 200, 400, 600, and 800 nm. For each value of $L_D$, the initial condition, Eq. (S7.2), is parameterized with the initial beam diameter $\sigma_0 = 455$ nm and initial exciton number taking values $N_0 = 1, 2 \ldots, 10$ (reflecting variation of the pump power). Subsequently, we attain widths of PL intensity profiles from the time-integrated signal $I(x) = \int_0^\infty n(x,t)dt$ by fitting it with the Gaussian function $I(x) = \langle n(0) \rangle \exp[-2x^2/\sigma^2]$ where $\langle n(0) \rangle = \int_0^T n(x=0,t)dt/T$ is the time averaged population at the center of the pulse ($T = 200$ ps). The results of the calculations are summarized in Fig. S8. For SWCNTs with $L_D \lesssim \sigma_0$ (blue and orange curves), an increase in the exciton density leads to a rise of PL profile width, $\sigma$, which is consistent with the SWCNT in Fig. 1a-1f of the main text. In contrast, $\sigma$ shows an initial drop followed by further recovery for $L_D > \sigma_0$ (green and red curves). The initial drop in the PL profile widths observed in the simulation is consistent with the observations in Fig. 1g-1l of the main text.

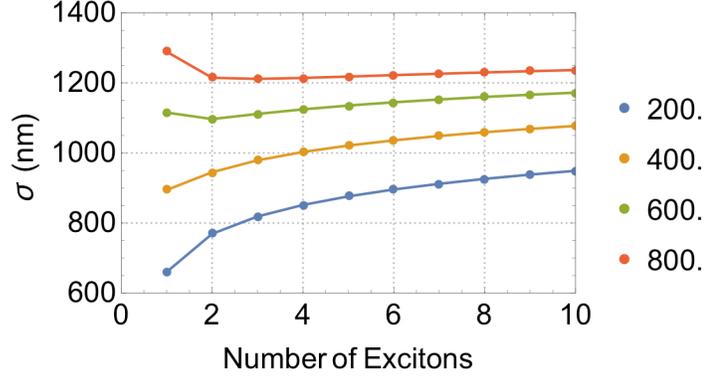

**FIG. S8:** Excitation number (i.e., pump power) dependent PL intensity profile width, $\sigma$, evaluated for 4 μm long SWCNTs with $L_D$ = 200, 400, 600, and 800 nm (color codes are in the legend). For comparison, excitation pulse diameter is set to $\sigma_0 = 455$ nm.

### S9 Auger recombination model

Consider excitons hopping between SWCNT segments associated with exciton mean free-path $\xi$. For a nanotube of length $L$, the probability to find an exciton at any *particular* segment is small, equal to $\xi/L$. For $N$ excitons distributed homogeneously along SWCNT, the probability to find any *two* excitons at one particular segment is given by the quadratic term of the binomial expansion of $(\xi/L + (1-\xi/L))^N$, namely

$$\frac{N(N-1)}{2}(1-\frac{\xi}{L})^{N-2}\left(\frac{\xi}{L}\right)^2 \simeq \frac{N(N-1)}{2}\left(\frac{\xi}{L}\right)^2. \quad (S9.1)$$

Multiplying this term by the number of segments, $L/\xi$, provides the probability of finding two excitons on any site of SWCNT,

$$P_N^{2X} = \frac{N(N-1)}{2}\left(\frac{\xi}{L}\right). \quad (S9.2)$$

Introducing an Auger recombination rate $\gamma_A$ associated with *a segment*, one can write

$$\frac{dN}{dt} = -\gamma_A P_{2X}^N = -\gamma_A \frac{N(N-1)}{2}\left(\frac{\xi}{L}\right). \quad (S9.3)$$

Rewriting this equation in terms of the exciton density $n = N/L$ results in

$$\frac{dn}{dt} \simeq -\gamma_A \xi \frac{n^2}{2} = -C_A n^2, \quad (S9.4)$$

allowing us to identify the Auger coefficient in terms of the Auger rate as

$$C_A = \frac{\gamma_A \xi}{2}. \quad (S9.5)$$

Eq. (9.5) is the central result connecting the Auger rate and the mean free-path. Identification of the former quantity depends on the model. To explain the linear dependence of $C_A \sim \gamma_A \xi / 2$ required by the fit of experimental data in Fig. 2 of the main text, we consider the following two models for the total Auger recombination process.

**(i) Direct Fermi golden rule model.** By taking into account that the excitons experience coherent motion on the lengthscale $\xi$, the Fermi golden rule expression for Auger recombination rate of two tightly bound excitons in 1d can be written as [31]

$$\gamma_A = 2\frac{A}{\xi}, \qquad (S9.6)$$

where microscopic Auger constant

$$A = 128 \frac{\omega_{cv}}{k_{e0}} \left(\frac{\mu}{m_0}\right) \left(\frac{E_b}{E_g}\right)^3 \qquad (S9.7)$$

depends on the interband transition strength $\omega_{cv} = \langle p \rangle_{cv}^2 / \hbar m_0$, the ratio $\mu/m_0$ of exciton reduced mass to electron mass in vacuum, the dissociated conduction band electron wave vector $k_{e0} = \sqrt{2\mu(E_g - 2E_b)}/\hbar$, the band gap energy $E_g$, and the exciton binding energy $E_b$. Using the following (6,5) SWCNT parameters, $\mu/m_0 = 0.05$, $E_g = 1.615$ eV, $E_b = 370$ meV, and $\hbar\omega_{cv} = 1$ eV, [16,31] we estimate $A = 109$ nm/ps. Substitution of Eq. (S9.6) into Eq. (S9.5) allows us to identify $C_A = A = 109$ nm/ps. Although this estimated value is close to the intercept point obtained by fitting experimental data in Fig. 2 of the main text, this model predicts $C_A$ to be *independent* of $\xi$. However, the linear correlation between $C_A$ and $D$ shows that $C_A \sim \xi$.

**(ii) Sequential model and capture-limited recombination.** We assume that two excitons confined within a segment of length $\xi$ can form a bound-state resonance within a capture time $\tau_c$, which is a precursor to Auger annihilation. The subsequent lifetime,

$$\tau_{2X,A} = 1/k_{2X,A}, \qquad (S9.8)$$

of the resonance will not depend on $\xi$ but will instead scale with the local interaction lengths of the *bound* two-exciton state and should be associated with a native native Auger rate $k_{2X,A}$ This describes a *sequential* recombination process, where the effective rate scales as the reciprocal of the sum of the two times,

$$\gamma_A = \frac{1}{\tau_c + \tau_{2X,A}} = \frac{1}{\tau_c + 1/k_{2X,A}}. \qquad (S9.9)$$

Assuming that, $\tau_{2X,A} \gg \tau_c$, the native recombination rate is slower that the exciton capture, one obtains $\gamma_A \approx k_{2X,A}$. According to Eq. (S9.5) this limiting case provides Auger coefficient

$$C_A = \frac{k_{2X,A}\xi}{2}, \qquad (S9.10)$$

that is linear in $\xi$.

To estimate $k_{2X,A}$ in Eq. (S9.10), we recall that the diffusion constant, $D = v\xi$, depends on the exciton mean velocity $v = \sqrt{k_B T / m^*}$, where $k_B T$ is thermal energy and the ratio of exciton effective mass to electron mass in vacuum is $m^*/m_0 \sim 0.05$. As a result, at room temperature we have $v = 300$ nm/ps. According to Eqs. (S7.5) and (S9.10), $d_0 = D/C_A = 2v/k_{2X,A}$ resulting in $k_{2X,A} = 2v/d_0$, which provides a room temperature estimate for the Auger rate, $k_{2X,A} = 100$ ps$^{-1}$, and associated recombination time $\tau_{2X,A} = 10$ fs. Accordingly, the exciton capture is an ultrafast process with timescale limited to $\tau_c < 10$ fs. A simple estimate of the capture time can be performed assuming that two excitons moving with mean velocity $v$ within a segment of $\xi$ do not form a bound state unless they meet each other (i.e., no long-range exciton-exciton interaction is present). In this case the capture time is an average exciton meeting time, $\tau_c = 2\xi/v$. For the mean free-path values, $\xi \sim 4-15$ nm, found from the spread of the measured diffusion constants, we estimate that $\tau_c \sim 10-100$ fs. These values are in contradiction with the requirement that $\tau_c < 10$ fs. We suggest that significant reduction in the capture time can be achieved if long-range exciton-exciton Coulomb interactions facilitating formation of the bound states are present in the system.

## S10 Independent quantum emitter and coarse-grained EEA models

The justification for a simplified coarse-grained description hinges on representing a SWCNT as a set of independent emitters prepared by an optical excitation along a SWCNT with a diameter $\sigma_0$. Assuming that excitons separated by distance $\gtrsim L_D$ cannot participate in the EEA processes (assumed to be the only processes associated with the exciton-exciton correlation), we define the number of independent quantum emitters as

$$m = \frac{\sigma_0}{cL_D} . \qquad (S10.1)$$

Here, $cL_D$ is the size of an independent emitter with a scaling parameter $c \sim 1-2$ introduced to facilitate fitting of the experimental data.

Provided $cL_D \leq \sigma_0$, we can neglect time dependent diffusion effects within the independent quantum emitters and introduce the coarse-grained EEA *rate* (dimensionality ps$^{-1}$) as

$$\bar{C}_{\bar{n}_i} = \frac{\bar{n}_i(\bar{n}_i - 1)}{2} \frac{\bar{C}_{EEA}}{cL_D} , \qquad (S10.2)$$

where $\bar{n}_i$ is the average number of excitons per $i$-th $cL_D$ segment of a SWCNT within $\sigma_0$ and the time-average EEA coefficient

$$\bar{C}_{EEA} = \frac{1}{T}\int_0^T C_{EEA}(t) dt \qquad (S10.3)$$

reflects the time-integrated nature of the measured PL intensity and degree of photon antibunching. Here, $C_{EEA}(t)$ is the EEA rate entering Eq. (1) and defined by Eq. (2) in the main text. The integration cutoff time is chosen to be larger than the exciton decay time, i.e., $T \gg 1/k_{1X}$. The scaling of the decay rates in Eq. (S10.2) with the number of excitons follows from an assumption that all multi-exciton recombination pathways have identical contribution to the considered decay processes.[32]

In the case of a single quantum emitter and narrow laser spot, (i.e., $\sigma_0 \leq cL_D$) the diffusion processes become important in defining the PL profile. Therefore, at the higher exciton densities the mean-field Eqs. (1) and (2) presented in the main text must be solved. Interpretation of the 2$^{nd}$ order photon correlation, $g^{(2)}$, experimentally limited to the two-exciton states on this lengthscale, requires solution of the coupled equations of motion for single- and two-exciton populations accounting for the EEA processes, and subsequent evaluation of the exciton and two-exciton correlation functions contributing to $g^{(2)}$. The analysis goes beyond the scope of this paper and will not be considered below.

In the opposite case when $\sigma_0 \geq L_D$, to estimate average number of excitons per $cL_D$ segment depending on the laser spot size $\sigma_0$, we assume a square beam profile (rather than Gaussian), which ensures uniform distribution of excitons within the $\sigma_0$ segment of SWCNT. Since a Poisson prefactor favors only two exciton states (Fig. S10b), we introduce a probability distribution for one exciton to be found within an interval $dx$ a distance $x$ from the other exciton (located at $x=0$). For a uniform exciton distribution this conditional probability distribution reads $P(x)dx = 2dx/\sigma_0$, where the prefactor 2 reflects the fact that the second exciton has equal probability to be found at coordinates $x$ and $-x$. According to this distribution, the probability of the second exciton to be within a distance not exceeding $cL_D/2$ from the other exciton (i.e., to belong to the same quantum emitter) is $P_2 = \int_0^{cL_D/2} P(x)dx = cL_D/\sigma_0$. The probability to be outside this interval (i.e., to belong to another quantum emitter) is $P_1 = 1 - P_2 = 1 - cL_D/\sigma_0$. Accordingly, the average number of excitons per quantum emitter can be estimated as

$$\bar{n} = 1P_1 + 2P_2 = 1 + \frac{cL_D}{\sigma_0} . \tag{S10.4}$$

Notice that the average number of excitons per independent emitter does not depend on the emitter number $i$, indicating their uniform distribution. We thus dropped this index from $\bar{n}$. Provided, $\sigma_0 = cL_D$ (single emitter case), $\bar{n} = 2$, showing that two excitons prepared by the excitation pulse are confined within a single emitter. If $\sigma_0 \gg cL_D$ (very large number of quantum emitters) then $\bar{n} = 1$ meaning that the two excitons have vanishing probability to meet within a single emitter.

The quantum yield for the transition from the state of two excitons to one exciton reads

$$Q_{2X} = \frac{\bar{n} k_{1X,R}}{k_{2X}} . \tag{S10.5}$$

Here, $k_{2X} = \bar{n} k_{1X} + \bar{C}_{\bar{n}}$ and the course-grained EEA constant, $\bar{C}_{\bar{n}}$ is given by Eqs. (S10.2) and (S10.3) with the average number of excitons defined by Eq. (S10.4). Finally, introducing the exciton quantum yield as

$$Q_{1X} = \frac{k_{1X,R}}{k_{1X}} , \tag{S10.6}$$

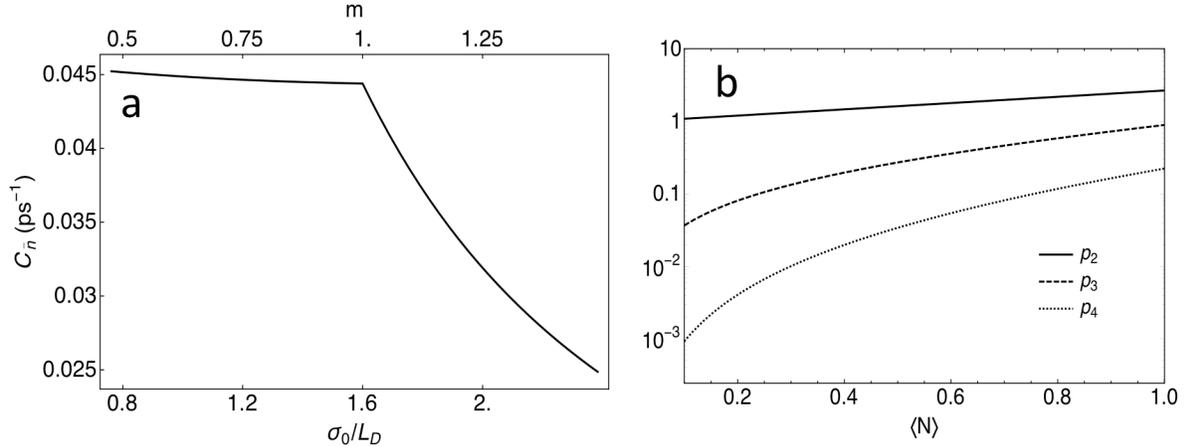

**FIG. S10 a:** Coarse-grained EEA rate given by Eq. (S10.2) with average number of excitons per quantum emitter given by Eq. (S10.4) calculated as a function of the ratio $\sigma_0/L_D$ and number of independent emitters $m = \sigma_0/1.6L_D$. **b:** The ratio of Poisson probabilities, $p_n = P_n^{\langle N \rangle}/\left(P_1^{\langle N \rangle}\right)^2$ to produce $n$ and one excitons, respectively, plotted on a log scale as a function of the average number of absorbed photons $\langle N \rangle$.

where $k_{1X} = k_{1X,R} + k_{1X,NR}$ is the exciton decay rate representing a sum of radiative and non-radiative decay rates, respectively. The ratio of the quantum yields entering an expression for the minimal degree of photon antibunching (Eq.3 in the main text) is

$$\frac{Q_{2X}}{Q_{1X}} = \left(1 + \frac{\bar{n}-1}{2}\frac{\bar{C}_{EEA}}{k_{1X}}\right)^{-1}. \quad \text{(S10.7)}$$

As described in the main text Fig. 4, Eq. (S10.7) shows a weak dependence on $\sigma_0/L_D$ for a single quantum emitter (i.e., $m \leq 1$). This is a consequence of a similar weak dependence of the coarse-grained EEA rate given by Eqs. (S10.2), (S10.3) with (S10.4) as demonstrated in Fig. S10a. In fact, in performing course graining, one should not expect that an averaged quantity will vary on a lengthscale below the size of a single independent emitter. This is the limitation of the adopted model discussed above.

The analysis of the measured 2$^{nd}$ order photon correlation function is based on the assumption that the number of photons, $\langle N \rangle$, absorbed by each SWCNT per pulse, is low. Accordingly the minimal photon antibunching ratio, $R_0$, is evaluated in the limit $\langle N \rangle \to 0$. [33] However, in the experiment discussed in the Letter (Fig. 3), the minimum average number of absorbed photons per pulse is estimated to be $\langle N \rangle \sim 0.2 - 0.6$. To account for the effect of finite $\langle N \rangle$, we consider a ratio $p_n = P_n^{\langle N \rangle}/\left(P_1^{\langle N \rangle}\right)^2$ of Poisson distribution functions, $P_n^{\langle N \rangle} = \langle N \rangle^n e^{-\langle N \rangle}/n!$. The prefactor $p_2$, accounting for the simultaneous absorption of two photons, is explicitly present in the main text Eq. (3). To rule out higher order photon absorptions processes that in principle could contribute to the 2$^{nd}$ order photon correlation function, we plot in Fig. S10b the ratios $p_3$ and $p_4$, and compare them with $p_2$. The plot clearly shows that simultaneous three and four photon absorption processes have negligible probabilities compared to the two-photon processes within the experimental range of $\langle N \rangle$.